\relax
\documentclass[letterpaper,twocolumn,10pt]{article}
\usepackage{times}
\usepackage{helvet}
\usepackage{courier}
\usepackage{url}
\usepackage{graphicx}
\usepackage{epigraph}
\usepackage[normal]{threeparttable}
\usepackage{graphicx}
\usepackage{bbding}
\usepackage{natbib}
\usepackage[letterpaper,left=0.75in,right=0.75in,top=0.75in,bottom=1.25in]{geometry}
\usepackage{titlesec}

\newcommand{\beq}{\begin{eqnarray}}
\newcommand{\eeq}{\end{eqnarray}}
\newcommand{\+}{\phantom{-}}

\titleformat*{\section}{\large\bfseries}
\titleformat*{\subsection}{\normalsize\bfseries}
\titleformat*{\subsubsection}{\normalsize\bfseries}
\titleformat*{\paragraph}{\normalsize\bfseries}
\titleformat*{\subparagraph}{\normalsize\bfseries}
\titlespacing\section{0pt}{12pt plus 4pt minus 2pt}{0pt plus 2pt minus 2pt}
\titlespacing\subsection{0pt}{12pt plus 4pt minus 2pt}{0pt plus 2pt minus 2pt}

\begin{document}
\title{Network Structure, Efficiency, and Performance in WikiProjects}
\author{
Edward L. Platt \\
University of Michigan\\
elplatt@umich.edu
\and
Daniel M. Romero\\
University of Michigan\\
drom@umich.edu
}
\date{}
\maketitle

\begin{abstract}
\small
The internet has enabled collaborations at a scale never before possible,
but the best practices for organizing such large collaborations are still not clear.
Wikipedia is a visible and successful example of such a collaboration which might offer
insight into what makes large-scale, decentralized collaborations successful.
We analyze the relationship between the structural properties of WikiProject coeditor networks
and the performance and efficiency of those projects.
We confirm the existence of an overall performance-efficiency trade-off,
while observing that some projects are higher than others in both performance
and efficiency,
suggesting the existence factors correlating positively with both.
Namely, we find an association between low-degree coeditor networks
and both high performance and high efficiency.
We also confirm results seen in previous numerical and small-scale lab studies:
higher performance with less skewed node distributions,
and higher performance with shorter path lengths.
We use agent-based models to explore possible mechanisms for
degree-dependent performance and efficiency.
We present a novel local-majority learning strategy designed to satisfy properties
of real-world collaborations.
The local-majority strategy as well as a localized conformity-based strategy
both show degree-dependent performance and efficiency,
but in opposite directions,
suggesting that these factors depend on both network structure and learning strategy.
Our results suggest 
possible benefits to decentralized collaborations made of smaller,
more tightly-knit teams,
and that these benefits may be modulated by the particular learning strategies
in use.
\end{abstract}

\maketitle

\section{Introduction}
\epigraph
{The problem with Wikipedia is that it only works in practice. In theory, it's a total disaster.}
{---Gareth Owen \cite{elsharbaty_editing_2016} }

The internet has enabled collaborations at a global scale.
Wikipedia, a free encyclopedia that invites anyone to edit articles,
is one of the most successful and visible examples of such a collaboration.
Organizing groups without top-down control is notoriously
difficult
\cite{freeman_tyranny_1972},
and yet Wikipedia, with millions of self-organized editors,
has produced a high-quality encyclopedia \cite{giles_internet_2005,keegan_evolution_2017}.
A better theoretical understanding of projects like Wikipedia is highly desirable as it could
help inform the design of new collaborative projects.
We focus on one aspect of a large-scale decentralized collaboration:
its network structure \cite{newman_structure_2003}.
How does Wikipedia's non-hierarchical structure relate to its success?

We look at WikiProjects on the English-language Wikipedia.
WikiProjects are collections of thematically related articles,
each with their own standards and norms.
When measuring the quality of collaborative projects,
there are at least two distinct measures to consider.
The first measure is short-term:
how effective a unit of work is at improving
the collaboration's output,
which we call {\em efficiency}.
The other measure is long-term:
the highest quality typically reached by an output,
which we call {\em performance}.
These two terms are often used interchangeably,
but we find it fruitful to distinguish between the two.
We find that Wikipedia exhibits an overall trade-off between performance and efficiency.
However, some WikiProjects surpass others in both efficiency and performance,
suggesting the existence of factors that correlate positively with both.

Our study focuses on the coeditor networks of each WikiProject:
which editors have edited at least one article in common?
These relationships represent the possible flow of information.
We focus specifically on mean degree, degree skewness, and path length.
High-degree editors have more collaborators,
which can increase diversity and access to information at the possible
expense of higher coordination costs
\cite{hong_groups_2004,golub_naive_2010}.
Highly skewed degree distributions can amplify the biases of high-degree
editors while reducing the need for explicit coordination
\cite{kearns_experiments_2012}.
Networks with shorter path lengths allow information to travel more quickly
at the possible expense of less local diversity
\cite{mason_propagation_2008,barkoczi_social_2016}.

In addition to our empirical study,
we use agent-based modeling to examine the consequences of specific
assumptions on networked collaboration.
We model individual behavior using a {\em social learning strategy} that
assumes agents 1. can only access a fraction of the model's state,
2. interact with others who share their concerns,
and 3. integrate their preferences into a single state.
Our model is the first we are aware of to incorporate these assumptions,
which are present across many real-world collaborations,
including Wikipedia.
\pagebreak

Our main findings are:
\begin{itemize}
\setlength\itemsep{0pt}
\item Despite an overall performance/efficiency trade-off,
WikiProjects with low-degree coeditor networks tend to have both higher performance and higher efficiency;
\item Short paths are associated with higher performance, consistent with a conformity-based learning strategy;
\item Structural inequality, as measured by degree skewness, is associated with lower performance;
\item Our agent-based model
shows that the efficiency and performance of collaborations can
depend on network degree,
and that the direction of that dependence varies with social learning strategy.
\end{itemize}

Our findings shed light on the importance of network structure for successful
collaboration.
These findings might be informative for future interventions that
recommend tasks based on how they will
influence network structure,
or for interventions that seek to encourage behaviors
complementary to existing network structure.

\section{Background and Related Work}
\label{sec:background}
The present paper investigates the relationship between social networks
and collaboration outcomes.
This connection has been explored by a number of theoretical,
numerical, and small-scale lab studies in the field of
{\em social learning}.
We contribute to this literature with a large-scale, empirical field study.
In much of the existing literature,
degree distribution correlates with outcome measures.
But aside from the naive Bayes case, it is unknown whether
the correlation is explained best by degree or by another structural property, such as characteristic path length.
In the empirical networks we study,
unlike artificial networks,
the structural properties vary independently,
making it easier to isolate individual
network properties that correlate with outcome variables.

{\bf Social Learning.} In {\em networked social learning},
agents are represented by nodes on a network
and can interact only with their neighbors.
Social learning tasks can be divided into cases where agents have {\em generated signals}
(independently noisy estimates of a true value)
and those where agents have {\em interpreted signals}
(solutions based on different selections of available data)
\cite{hong_interpreted_2009}.
The behavior of individual agents is described by their
{\em social learning strategy}.
For generated signals,
a naive Bayesian approach converges to the truth
when all agents have the same degree,
while the speed of convergence depends on the {\em spectral gap}
between the two largest eigenvalues of the network's interaction matrix
\cite{degroot_reaching_1974,golub_naive_2010}.
Complex social learning tasks can also be modeled as the problem
of maximizing an objective function with many local maxima,
referred to as a {\em rugged landscape}
\cite{lazer_network_2007,mason_propagation_2008,mason_collaborative_2012,grim_scientific_2013,barkoczi_social_2016}.
Numerical simulations have shown that efficient networks
(those with short paths between nodes)
can result in faster convergence at the cost of a less optimal solution,
due to less time for exploration
\cite{mason_propagation_2008,grim_scientific_2013}.
However, when conformity-based social learning strategies are used,
efficient networks can sometimes find more optimal solutions than
inefficient ones \cite{barkoczi_social_2016}.
Using an agent-based model, Hong and Page \cite{hong_groups_2004} found that
diverse groups can outperform groups composed of the best individual
problem-solvers.

{\bf Lab experiments.} Lab-based experiments on networked collaboration
suggest a complex interaction between network topology and other factors.
While groups of networked human subjects perform very well on
difficult graph-coloring tasks, the best performing network architectures
(e.g., fully-connected vs. small-world) vary
from task to task \cite{kearns_experiments_2012}.
The same studies found that while human subjects tend to perform well on
many networks, they perform worst on self-organized
networks, possibly due to higher structural inequality
(degree skewness).
Similarly, some network topologies are able to reach faster decisions in the
presence of more information, while others show the opposite effect
\cite{kearns_experimental_2006}.
Based on lab experiments, Fowler and Christakis \cite{fowler_cooperative_2010}
suggest that individual decisions towards altruism are conditional on their
neighbor's behavior and ``contagious'' up to three degrees away.
Later experiments by Suri and Watts \cite{suri_cooperation_2011} confirmed the
existence of conditional altruism,
but concluded that altruism
influences only first-degree neighbors.

{\bf Digital Communities.}
Research on digital communities has also examined the role of 
diversity and inequality
in collaborative work and decision-making.
In sociology, research has focused on the relationship between
network structure and social capital.
Powerful individuals are often ``brokers''
who act as exclusive intermediaries between disconnected portions of the
social network \cite{silverman_patronage_1965}.
Similarly, successful innovation in organizations often occurs in ``structural
holes'' between groups \cite{granovetter_strength_1973}.

For Wikipedia specifically,
Robert and Romero \cite{robert_crowd_2015} found that
larger group sizes yield higher article ratings
when the groups are diverse and experienced. Kittur and Kraut found that different types of coordination have a complex
effect on the quality of Wikipedia articles \cite{kittur_harnessing_2008}.
Both explicit and implicit coordination result in higher quality articles,
with explicit coordination being especially central in the early life of an
article.
Shaw and Hill \cite{shaw_laboratories_2014}
found that behavior in online wiki communities is consistent
with the ``iron law of oligarchy,''
which states that
earlier members of a group will, over time, gain disproportionate
decision-making power and act increasingly out of self-interest rather than
the good of the group \cite{michels_political_1999}.
Similarly, Halfaker et al. \cite{halfaker_rise_2013} attributed decreasing
participation on Wikipedia to poor retention of new users.
Looking specifically at Wikipedia policies determined by editor consensus,
Keegan and Fiesler \cite{keegan_evolution_2017} found a trend
from flexible rule-making towards less flexible maintenance and deliberation.
Using content analysis,
Morgan et al. \cite{morgan_project_2013} found WikiProjects to be
more loosely organized than traditional teams.

Across the broad range of work discussed above,
a few key themes emerge.
Both the efficiency and the performance of a collaboration are important
considerations and vary depending on both network structure and type of task
\cite{kearns_experiments_2012}.
While generated signal models of social learning predict no relationship
between the two
\cite{golub_naive_2010},
contagion-style innovation models predict a trade-off
\cite{mason_collaborative_2012,barkoczi_social_2016}.
Such a trade-off has been observed in simulations and lab experiments on
collaboration
\cite{kearns_experiments_2012,grim_scientific_2013}.

\section{WikiProjects}
\label{sec:wp}

\begin{figure}[t!]
\begin{center}
\includegraphics[width=3in]{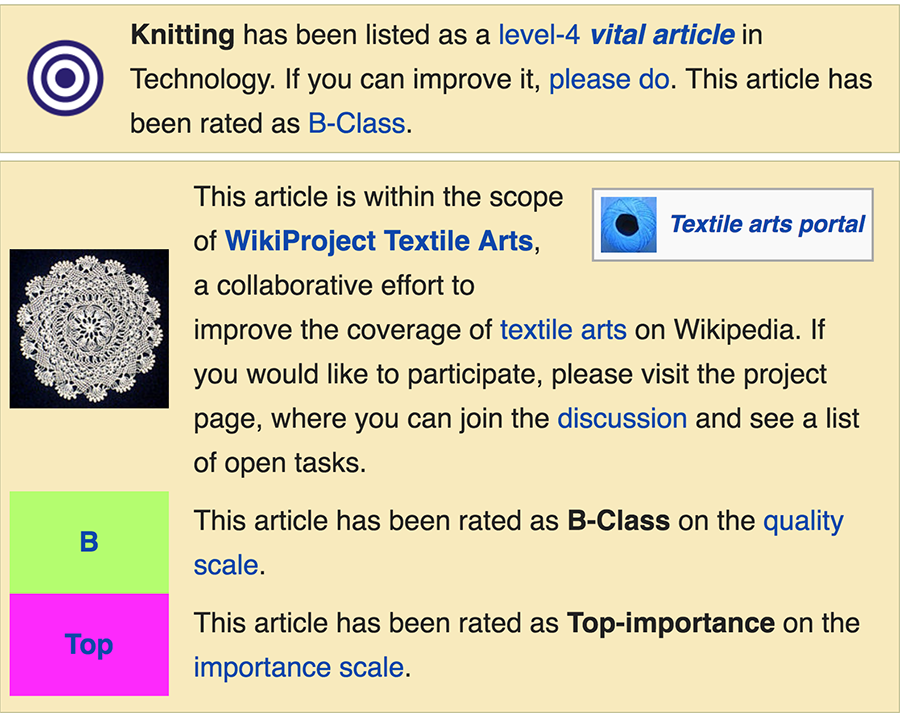}
\caption{
From Wikipedia {\em Knitting} talk page.
Two WikiProjects have assessed the article as B-class quality.
\label{fig:knitting}
}
\end{center}
\end{figure}

Many articles on Wikipedia belong to one or more WikiProjects.
WikiProjects are groups of thematically-related articles
(e.g., articles related to Philosophy).
Information about an article's associated WikiProjects
can be viewed on that
article's talk page (Figure \ref{fig:knitting}).
Each WikiProject has its own page and talk page,
containing information about conventions within the project
as well as discussions about individual articles.
WikiProjects are thus distinct communities, with distinct norms and processes.
These communities are the fundamental units of analysis in this paper.

One of the main roles of a WikiProject is to evaluate the quality of its articles.
Quality assessments are made through consensus-based deliberation on the WikiProject
talk page.
Within a WikiProject,
assessments are typically made using the following {\em assessment classes}
(in order of increasing quality):
Stub, Start, C, B, A.
Different WikiProjects can assign different quality assessments to the same
article.
Differences between quality assessments could reflect different quality standards,
different grading systems, different responsiveness to changes in an article, etc.

In addition to the above assessment classes, articles on Wikipedia can be tagged as
``good article'' (GA) or ``featured article'' (FA) quality.
FA and GA determinations are made using a Wikipedia-wide consensus,
independently of WikiProject-based evaluations.
FA articles are ``the best articles Wikipedia has to offer''
\cite{wikipedia_contributors_wikipedia:featured_2018}.
GA articles meet ``a core set of editorial standards`` but are ``not featured article quality''
\cite{wikipedia_contributors_wikipedia:good_2017}.
When an article is assigned GA or FA status,
WikiProject quality assessments are often updated to reflect that status.
For example, the article {\em Mewtwo} was assessed as GA status on October 5,
2009 and shortly afterwards
its quality assessment was changed from B to GA within both
{\em WikiProject Pok\'emon} and
{\em WikiProject Video Games}.
This example also illustrates a quirk of conventions on Wikipedia:
very often, articles pass to GA or FA directly from B, skipping A.
The majority of WikiProjects rarely use the A class quality assessment.

\subsection{Data}

Our analysis combines multiple data
sets from the English-language Wikipedia \cite{platt_english_2018}.
For information about edit history, we used a publicly-available data set containing
metadata (time, article id, user) about all edits from July 12, 2006 to December 2, 2015.
We used a custom script to scrape article quality assessments from logs produced
by WP 1.0 Bot for 2279 unique WikiProjects
between May 4, 2006 and December 2, 2015.
Finally, we used a publicly-available database dump of page events (including rename events)
to reconstruct the article id for each title mentioned in the assessment logs.

\subsection{Efficiency and Performance}

When individuals collaborate to solve a problem,
there are many ways to gauge their success.
One possibility is {\em efficiency}:
how quickly they find a solution.
Another is {\em performance}:
how good their solution is.
Evidence from numerical simulations
\cite{lazer_network_2007,mason_propagation_2008,mason_collaborative_2012,grim_scientific_2013,barkoczi_social_2016},
lab studies \cite{kearns_experiments_2012},
and field observations \cite{gentry_consensus_1982}
all suggest a trade-off between efficiency and performance.
While common, this trade-off is not absolute,
suggesting it is sometimes possible to simultaneously increase
performance and efficiency.
The identification of factors associated with both higher efficiency
and higher performance has obvious practical importance.
In this paper, we focus on how network structure
relates to efficiency and performance within WikiProjects.

For a WikiProject, efficiency quantifies how much participants can raise the
assessed quality of an article for a fixed amount of work.
We measure work by the number of revisions made.
Quality assessments are made through consensus of the project participants themselves.
Different projects can have different standards and practices for assessing article quality,
so the efficiency is not a measure of how quickly some objective measure of quality improves,
but rather of how quickly the project participants can reach consensus on the improvements that
need to be made and make those improvements.
Because our definition relies on assessment transitions,
we must define efficiency variables for
each of the project-level quality assessments: A, B, and C.
For a particular grade $G$,
we desire our definition of efficiency to meet the following conditions:
\begin{itemize}
\setlength\itemsep{0pt}
\item{Strictly increasing in the number of articles reaching grade $G$ (with revision count fixed);}
\item{Strictly decreasing in the number of revisions (with transition count fixed);}
\item{Independent of WikiProject size: not affected by adding an article having the same efficiency.}
\end{itemize}

We now define an efficiency measure which meets the above criteria.
Let $T(W,G)$ be the set of article assessment transitions from below grade $G$
to grade $G$ or higher in project $W$.
Let $N(W,G)$ be the number of articles in project $W$ which ever transition
from below grade $G$ to grade $G$ (or higher).
Given a transition $t$,
let $r(t)$ be the number of revisions to the article
since its previous grade transition,
and let $g(t)$ be the number of grade levels crossed bt $t$.
We quantify the efficiency $E(W,G)$ as the inverse of the mean number of revisions
per transition:
\beq
E(W,G)
&=&
\left[
\frac{1}{
N(W,G)
}
\sum_{t \in T(W,G)} \frac{r(t)}{g(t)}
\right]^{-1},
\eeq
where the $g(t)$ term accounts for assessments that raise article quality by
several grades by
dividing the revisions evenly between all grade levels achieved.

For performance, we wish to quantify how good articles tend to be when they reach a stable state.
Measuring performance is difficult for two reasons:
there is no objective measure of article quality available,
and articles are always changing, making it difficult to know which articles should be considered
complete or stable.
We use an extremely simple performance measure that gives surprisingly consistent results.
In addition to per-project quality assessment, articles can be given ``featured article'' or
``good article'' status.
The criteria for these statuses are consistent across all of Wikipedia,
and any editor can participate in the discussion and decision to award good or featured
status.
In other words, the good and featured statuses are less subjective than per-project assessments.

Our performance measure $P(W)$ is defined simply as
the percentage of articles in project $W$ which have reached
good or featured status:
\beq
P(W) &=& \frac{f(W) + g(W)}{n(W)},
\eeq
where $f(W)$ and $g(W)$ are the numbered of featured and good articles respectively,
and $n(W)$ is the total number of articles.

\subsection{Coeditor Networks}

We would like to determine how the social network structure of
Wikipedia---the pattern of who interacts with whom---relates to
efficiency and performance.
There are several types of interactions we could focus on,
including:
coediting, user talk messages, and talk page replies.
We choose to focus on coediting: when two editors have made changes to the
same article or talk page.
While editors can communicate directly through user talk messages,
the number of such messages is small compared to the number of edits to article
and talk pages.
We also could have considered direct replies between editors on article talk
pages, but these replies are typically seen (and intended to be seen)
by everyone reading the talk page,
and are part of larger conversations.
When an editor views a page,
they are potentially viewing content from and interactions
between all editors who came before them,
motivating our choice to focus on the social network structure of
coeditors.

The {\em coeditor network} of a WikiProject consists of nodes representing editors
and edges connecting pairs of editors who have edited the same article.
The edges are directed, with the direction representing
{\em plausible information flow};
an edge from Alice to Bob exists if Alice edited an article and then Bob edited the same article at
a later time. Note that edges can exist in both directions. 
We make the simplifying assumption of unit weight for all edges.
We focus on three structural properties:
degree, characteristic path length, and min-cut.
Degree and characteristic path length have been shown to correlate with
performance and efficiency in some social learning settings
\cite{golub_naive_2010,mason_propagation_2008,grim_scientific_2013},
while min-cut can be interpreted as a measure of decentralization,
common feature of peer-produced communities such as Wikipedia
\cite{benkler_wealth_2006}.

The degree distribution is the simplest network property we analyze.
The in-degree (out-degree) of a node is the number of edges to (from) that node.
Taking the average of either in-degree or out-degree gives the same value:
the {\em mean degree} of the network.
In our context, the mean degree represents, on average,
how many other editors each editor has collaborated with.
We also consider the {\em skewness} of the in-degree and out-degree distributions.
A large positive degree skewness for a WikiProject coeditor network
implies that a small number of editors have a very large number of collaborators,
while a small positive value implies that the editors having the most collaborators
don't have many more than a typical editor.

We also calculate the characteristic path length for each WikiProject coeditor network.
The {\em distance} from node $s$ to node $t$ is the distance of the shortest path
from $s$ to $t$.
The {\em characteristic path length} (or just {\em path length})
is the mean distance between all editor pairs,
excluding unconnected pairs.
To account for unconnected nodes, we also measure the {\em connected fraction}:
the fraction of ordered node pairs with a directed path from source to sink.
The path length represents how quickly information can move through the network.
Networks with longer paths require more interactions for information to propagate,
which has been shown to reduce efficiency in some settings
\cite{mason_propagation_2008,barkoczi_social_2016}.

Our final network measure quantifies the connectivity of a project's coeditor
network using min-cut size.
The minimum $st$-cut between nodes $s$ and $t$ is the set of edges that must beremoved
for no path exists from $s$ to $t$.
The minimum cut (min-cut) of a graph is the smallest minimum $st$-cut over all node pairs $st$. 
The size of the graph min-cut quantifies the connectivity of a graph,
but only incorporates information about edges lying on paths crossing the min-cut.
Instead, we use the mean size of all minimum $st$-cuts, which we refer to as the
{\em mean min-cut}.
This measure quantifies the number of redundant paths information can take through the network.
Networks with higher redundancy are more resilient to errors on one path \cite{albert_error_2000}
and allow innovations to propagate through complex contagion,
in which innovations are only adopted after multiple exposures \cite{centola_complex_2007}.

The mean path and min-cut are computationally intensive,
requiring distance and minimum $st$-cut calculations for all node pairs.
For larger projects, these calculations are impractical and we thus employed
sampling to determine mean path length and mean min-cut.
For mean path length, source nodes were sampled, and path length was calculated to all destination nodes
from each of these.
For min-cut, node pairs were sampled.
In both cases, stratification was used to ensure the same number of nodes were were sampled from each of
12 node degree quantiles.
We estimated the error due to sampling by determining true values for a medium-sized project,
and calculating error as a function of sample-size.
Sample sizes were chosen such that relative error was below 10\%.
Even with sampling, however, it was impractical to calculate these properties for the largest projects,
so we exclude the 183 largest projects from the analysis.

\begin{figure}[t!]
\centering
\includegraphics[width=2.5in,height=2.5in]{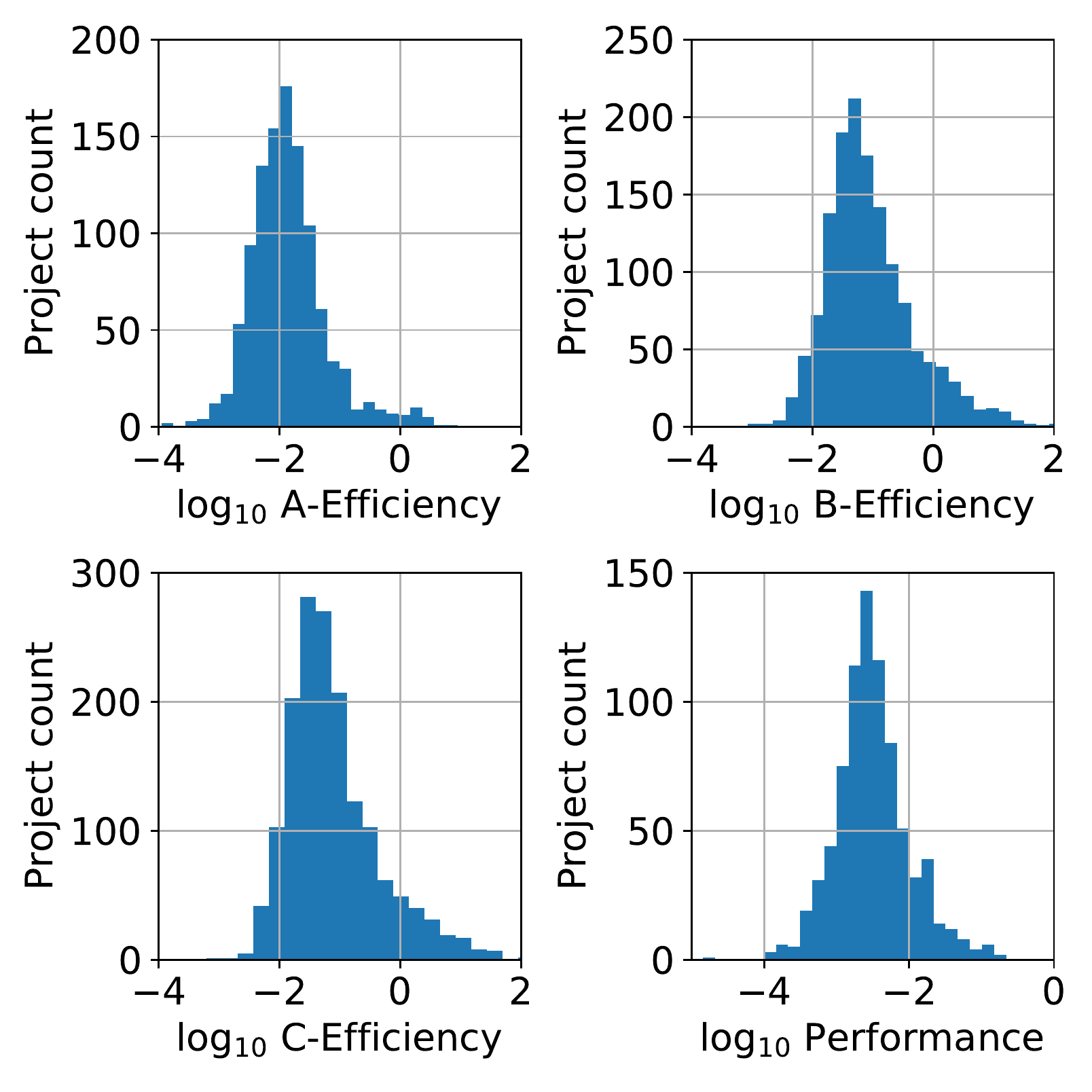}
\caption{
Histograms of WikiProject efficiency and performance.
Both measures are highly right-skewed, but form unimodal distributions
with low skewness after log transformation.
\label{fig:eff-perf-hist}
}
\end{figure}

\subsection{Empirical Results}

\begin{figure}[t!]
\centering
\includegraphics[width=2.5in,height=1.67in]{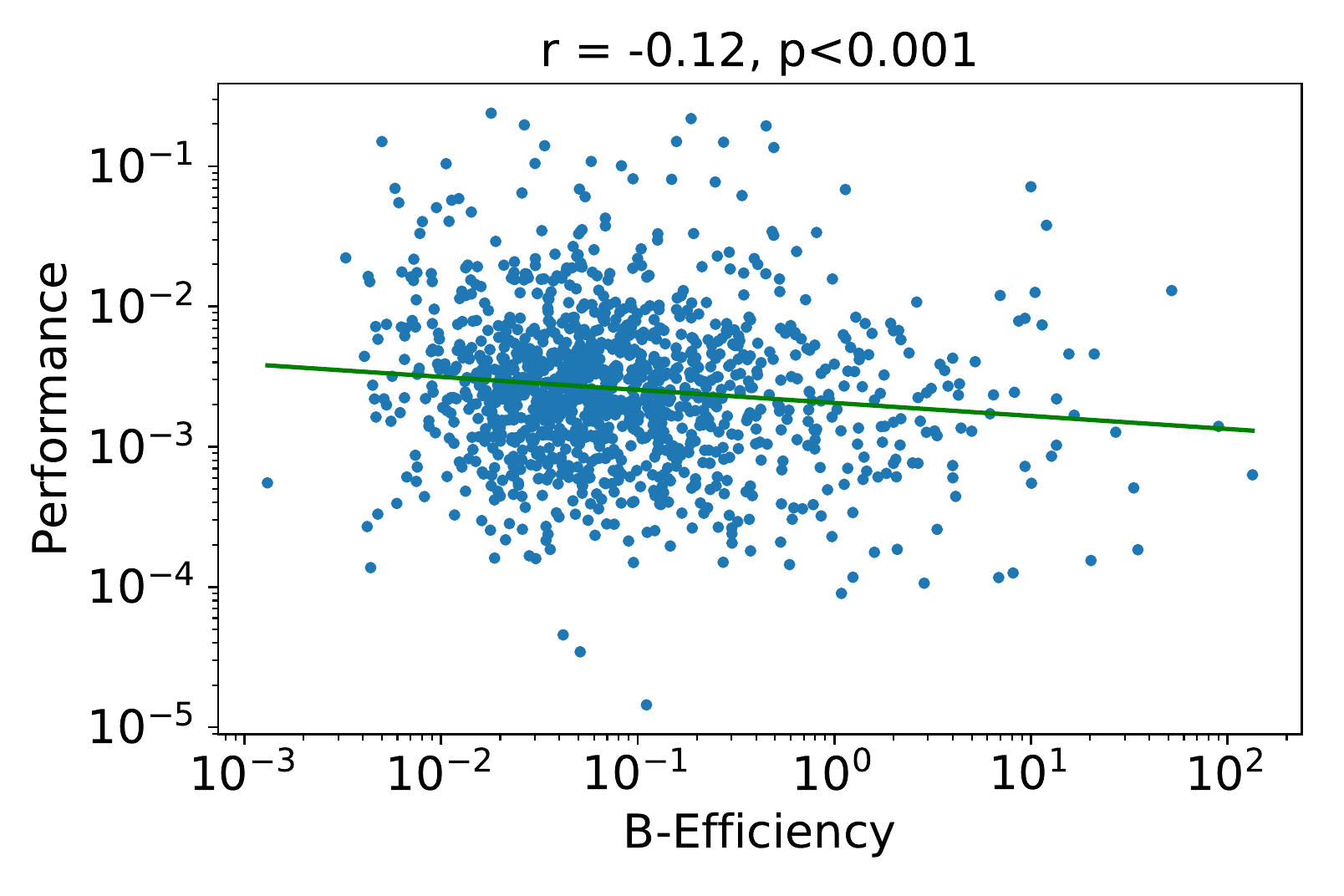}
\caption{
WikiProject performance is anticorrelated with B-level efficiency,
with Pearson r of -0.12.
Results are similar for other grade levels.
On average, highly efficient WikiProjects are under-performing,
but when looking at specific WikiProjects,
some are higher than others in both performance and efficiency.
\label{fig:eff-perf}
}
\end{figure}

We find that both efficiency and performance are highly right-skewed,
with a small number of projects having values much higher than the average.
After log-transforming the values, both the efficiency and the performance have
a unimodal distribution with low skew (see Figure \ref{fig:eff-perf-hist}).
Our findings confirm the trade-off between performance and efficiency
observed in many other settings (Figure \ref{fig:eff-perf}).
However, when looking at specific projects, some are higher in both performance
and efficiency,
suggesting the existence of factors which correlate positively with both.

We also find that mean min-cut is highly correlated with degree ($r=0.980$, $p<0.001$),
so we exclude min-cut from regression models to prevent collinearity.
The high correlation between mean degree and min-cut implies that,
in most cases,
the minimum $st$-cut is simply the set of edges from $s$
or the set of edges to $t$.
The rarity of non-trivial min-cuts suggests that WikiProject coeditor
networks have very few central
bottlenecks and are thus highly decentralized.

\begin{table}
\small
\begin{tabular}{lllllllll}
&Perf$^\dagger$&A-Eff$^\dagger$&B-Eff$^\dagger$&C-Eff$^\dagger$\\
\hline
Mean degree$^\dagger$&-0.7$^{***}$&-0.8$^{***}$&-0.6$^{***}$&-0.3$^{*}$\\
Out degree skew$^\dagger$&-0.4$^{***}$&-0.5$^{**}$&-0.3$^{*}$&-0.06\\
Mean path length$^\dagger$&-0.33$^{***}$&-0.09&-0.05&-0.09\\
C-Efficiency$^\dagger$&-0.08$^{*}$&\+---&\+---&\+---\\
Connected frac.&0.01&0.09$^{*}$&0.15$^{***}$&0.06\\
Talk fraction$^\dagger$&0&-0.02&-0.03&0.01\\
Mean similarity$^\dagger$&0.06$^{**}$&-0.03&0.01&0.02\\
Mean editors/art.$^\dagger$&0.3$^{**}$&0.3&0.2$^{*}$&0.09\\
Article count$^\dagger$&-0.4&0.7$^{*}$&0.8$^{**}$&0.7$^{**}$\\
Editor count$^\dagger$&0.4&0.9$^{**}$&0.8$^{**}$&0.5$^{*}$\\
Revision count$^\dagger$&0.6$^{*}$&-1$^{**}$&-1.1$^{***}$&-1$^{***}$\\
First assessment&0.05&0.11$^{**}$&0.31$^{***}$&0.43$^{***}$\\
Mean article age&-0.03&-0.04&-0.01&-0.05$^{*}$\\
\hline
N&1179&966&1260&1415\\
R$^2_{adj}$&0.37&0.17&0.30&0.43\\
\hline
\end{tabular}
\begin{tablenotes}
\item $\dagger$ Log-transformed. * $p < 0.05$. ** $p < 0.01$. *** $p < 0.001$.
\end{tablenotes}
\caption{Standardized coefficients for OLS models.
\label{tab:model}
}
\end{table}

To study the relationship between network structure, efficiency, and performance,
we model the performance and efficiency of WikiProjects using ordinary least-squares linear regression.
Each WikiProject is a single observation.
The models include each project's coeditor network properties as independent variables.
We also include the following project-level variables to control for confounding factors.
\begin{description}
\setlength\itemsep{0pt}
\item[C-efficiency]
(performance only).
Quantifies how quickly a WikiProject improves articles.
Efficiencies for different grades are highly correlated,
so we include only one.
\item[Connected fraction.]
Fraction of coeditor pairs connected by a path.
\item[Talk fraction.] Fraction of total revisions made to talk pages.
\item[Mean similarity.] Mean Jaccard similarity (by article) with other WikiProjects; a measure of topical complexity.
\item[Mean editors/article.] Mean number of editors collaborating on each article in a WikiProject.
\item[Article count.] Total number of articles in the WikiProject.
\item[Editor count.] Total number of editors working on articles within a WikiProject.
\item[Revision count.] Total number of revisions to articles in a WikiProject.
\item[First assessment.] Timestamp of first assessment; a measure of how long a WikiProject has been active.
\item[Mean article age.] Mean age of articles within a WikiProject.
\end{description}

Our models are summarized in Table \ref{tab:model}.
Min-cut is excluded from all models to avoid collinearity,
as it is highly correlated with degree.
In-degree and out-degree skewness were also highly correlated,
so we only include out-degree skewness
(results are similar for in-degree skewness).
Heavy-tailed variables are log-transformed.
To test the robustness of our results,
we also computed models using cube root instead of logarithmic transformations,
and using only top- and high-importance articles.
The results were qualitatively similar results for all variables,
except for degree-skewness, which had an inconsistent sign across models.

We see that B-efficiency and C-efficiency have very similar models, but that A-efficiency behaves
differently in its dependence on degree skewness and connectivity.
The different behavior of A-efficiency is likely explained by the
observation that the A-Class quality is infrequently used in practice.
The A-Class quality level is usually passed when an article reaches
good or featured article status,
which follow deifferent a consensus process from other ratings.

The negative dependence of performance on C-efficiency suggests there is generally a trade-off between
performance and efficiency.
However, low degree is correlated with both higher efficiency and higher performance,
suggesting that it is possible to improve both simultaneously.
Much of the existing numerical work on networked social learning focuses on path length rather than degree,
so we explore this result further using simulations in the next section.

For path length, we find that longer lengths correspond to lower performance, contrary to the conjecture
that longer path lengths allow more exploration \cite{mason_propagation_2008}
but consistent with a conformity-based social learning strategy \cite{barkoczi_social_2016}.

We also observe that high degree skewness is correlated with lower performance and lower A-efficiency,
suggesting that articles in projects with decentralized coeditor networks reach featured or good status
more efficiently, and reach higher quality ratings in general.

\section{Agent-Based Model}
\label{sec:sim}

\begin{table*}[!ht]
\small
\centering
\begin{tabular}{lccccc}
Name          & Social stage & Individual stage & Limited concern & Unknown objective & Single truth \\
\hline
Best+I & Best neighbor   & Global & & & \\
Conf+I & Conformity      & Global & & \Checkmark & \\
Best+LI & Best neighbor  & Local  & \Checkmark & & \\
Conf+LI & Conformity     & Local  & \Checkmark & \Checkmark & \\
LMaj+LI & Local majority & Local  & \Checkmark & \Checkmark & \Checkmark \\
\hline
\end{tabular}
\caption{
Definitions and properties of social learning strategies.
Each consists of a social stage and an individual stage.
Individual stages use hill-climbing based on either the global state,
or the agent's local concern.
\label{tab:strat}
}
\end{table*}

In addition to our empirical study,
we use a simple agent-based model of collaboration to better understand
the relationship between node degree, efficiency, and performance.
Numerical models allow us to determine the effect of changing a single
variable (e.g., network structure, learning strategy),
which is impractical in the empirical setting.
It is important to note that the goal of our model is not to simulate
all the intricacies of Wikipedia or any other specific platform.
Rather, our goal is to determine whether the correlations we observe
between degree and outcome variables
on Wikipedia can be reproduced in a more general setting.

Past work in the field of social learning typically models collaboration
as an optimization problem:
finding a state of the world which maximizes some objective function
\cite{lazer_network_2007,mason_propagation_2008,mason_collaborative_2012,barkoczi_social_2016}.
Wikipedia itself can be regarded as an optimization problem.
On Wikipedia, editors are generally seeking to improve the quality of articles
and have some personal preference over possible states of an article.
When editors do not agree on the optimal state of an article,
the conflict is resolved through a consensus-based deliberation.
This consensus process can be regarded as a {\em social choice function}
\cite{arrow_social_2012,brandt_computational_2012}
which maps individual preferences to community preferences.
Wikipedia can thus be thought of as a group of editors with individual preferences
for article states,
collaborating to optimize articles according to community preferences.
Note that these community preferences do not assume the existence of any
ground truth, other than the preferences themselves.

To simulate collaboration, we need a model problem for collaborators to solve.
Following existing literature on social learning,
we use the NK model \cite{kauffman_towards_1987}
to create NP-hard, nonlinear optimization problems.
The NK model produces an objective function with a
{\em rugged landscape}, i.e., many local optima.
The ruggedness of the model can be tuned through the parameters $N$
(the dimensionality of the solution space)
and $K$ (the level of inderdependence between dimensions).
Formally, the NK model produces an objective function $F$ mapping a binary string $S$ of length
$N$ to a real value in $[0,1]$.
Model state is divided into $N$ {\em loci}, with locus $i$ having a binary state $S_i$
and a value $f_i(S)$ dependent on its own state
and on the state of $K$ random other loci.
The functions $f_i(S)$ are created by selecting a random value in $[0,1]$ for
each possible state of locus $i$ and its $K$ neighbors.
The value of the model $F(S)$ is the mean of all locus values $f_i(S)$.
In our simulations,
agents iteratively search for a bit string $S$ that maximizes $F(S)$.

In a typical social learning model,
a set of agents each maintain an estimate of the optimal state
and iteratively update that estimate based on information available from other agents,
according to some {\em learning strategy}.
In networked social learning,
agents are associated with the nodes of a network and share information
only with their neighbors.
We define efficiency and performance in terms of the solution values for each time step
(averaged over many trials).
We define the performance to be the mean solution value after the process has converged,
while the efficiency is the reciprocal of the number of steps required to converge.
We measure the time to convergence as the number of steps required to reach 99\% of the maximum
mean solution value.

Without additional constraints,
the above model is missing
several key properties of real-world collaborations.
In designing our agent-based model, we paid attention to the
following properties.
\begin{description}
\setlength\itemsep{0pt}
\item[Limited concern.] Agents are concerned only with a subset of the entire
state when making decisions and determining preferences.
(On Wikipedia, editors typically interact with a small subset of the articles.)
\item[Concern-based network.] Agents interact with other agents who share a
common concern over some subset of the state.
(On Wikipedia, editors interact with others who share interests in the same articles.)
\item[Unknown objective.] Agents rank states in order of preference,
but do not have access to the objective function.
(On Wikipedia, there is no ground truth measure of quality.)
\item[Single source of truth.] At any given time, the system is in a single state
and agent preferences are based on local modifications to that state.
(At any point in time, there is only one current version of Wikipedia.)
\end{description}

\subsection{Concern-Based Networks}

On Wikipedia, editors interact by editing articles and talk pages.
Thus, the editors who interact with each other are exactly those who care about the same content.
Rather than using arbitrary networks,
we devise a network structure inspired by the above observation.
We do so by associating agents with particular loci in the NK model.
We also wish to study the effect of varying network degree,
which we achieve through a rewiring process described below.

Our concern-based networks are generated directly from the structure of the NK model.
The value of each NK locus depends on its own state and the state of $K$ other loci.
For each locus, we define an agent and assign these $K+1$ loci as its concern.
Next, an agent-agent co-affiliation network is created by connecting two agents if they share at least
one locus in their concerns.
This process is analogous to our construction of WikiProject coeditor networks.

To create a tunable degree, we duplicate each agent and its concern,
then randomly rewire a fraction of agent concerns before creating the agent-agent network.
With no rewiring, the duplication process creates a high overlap between agent concerns.
This overlap results in redundant links to a small number of agents,
rather than unique links to a large number of agents,
and therefore to an agent-agent network with small average degree.
By randomly rewiring the agent concerns, the redundancy is reduced
and the average degree of the agent-agent network is increased.

\subsection{Networked Learning Strategies}

Learning strategies determine how agents update their preferences based on
available information \cite{barkoczi_social_2016}.
Agents can engage in individual learning
by applying a hill-climbing algorithm to their current solution.
In each iteration, one bit of the NK solution string is flipped to maximize the solution value.
If no change improves the value, the original solution is kept.
The above strategy relies only on rankings of states,
satisfying the unknown objective assumption.
However, it relies on information about the entire state,
violating the limited concern assumption.
In order to satisfy this assumption,
we also define a local variant
in which only a subset of bits in the NK solution string are considered.
This variant reflects a more realistic style of collaboration,
in which individual agents focus on sub-problems.

In social learning,
agents can also incorporate information from
other agents they are connected to by an edge.
While individual learning always converges to the local maximum relative to the starting point,
social learning strategies allow agents to ``jump'' to drastically different solutions with higher local maxima.
In our model, we use both the conformity and best-neighbor strategies from \cite{barkoczi_social_2016}.
In the {\em best-neighbor} strategy, each agent compares its solution to a sample of neighbors, and chooses the solution with the highest value.
In order to compare solutions between neighbors,
the exact value of the objective function must be known for each solution,
so this strategy does not satisfy the unknown objective assumption
or the limited concern assumption.
In the {\em conformity} strategy, agents simply choose the most common solution among their neighbors
(ties are broken uniformly at random).
This strategy does not rely on solution value at all, so clearly satisfies the
unknown objective and limited concern assumptions.
In both cases, a single iteration of individual learning is performed after each social learning
iteration.
Because each agent maintains a separate estimate of the solution,
neither strategy satisfies the single source of truth assumption.

\subsection{Local Majority Strategy}

To satisfy the single source of truth assumption,
we introduce a new strategy: {\em local majority}.
In local majority, agents all begin with the same starting state and apply individual learning to their concern to generate
possible improvements to the solution.
Next, a new solution is constructed by considering each locus of the NK solution
individually.
Every agent concerned with a locus votes for its state based on their preferred
new solution and the majority state is chosen.
The result of this process is that all agents integrate their solutions into
a single state,
which forms the basis for the next iteration.
This strategy more realistically reflects collaborations like Wikipedia:
at any given time, a Wikipedia article has a single state, determined by consensus,
but editors may have differing opinions on how to improve that article.

\subsection{Simulation results}

We simulated 100 trials for rewiring values of 0.0, 0.167, 0.333, 0.5, 0.667, 0.833, and 1.0.
For each trial we generated an NK model with $N=250$ and $K=7$,
generated a concern-based network, and ran each social learning strategy
(Table \ref{tab:strat}) for 300 iterations.
For conformity and best-neighbor strategy, we used a sample size of 3, following
\cite{barkoczi_social_2016}.
We confirmed that all trials converged to their maximum value before reaching
the last iteration.
Networks had mean degree 116.6 with 1.3 standard deviation,
and mean path length of 1.766 with 0.0027 standard deviation.
The coefficient of variation for degree is approximately 10\%,
while only 1\% for mean path length, confirming that the rewiring process has a stronger influence
on degree than on path length.

\begin{figure}[t!]
\centering
\includegraphics[width=3in,height=1.5in]{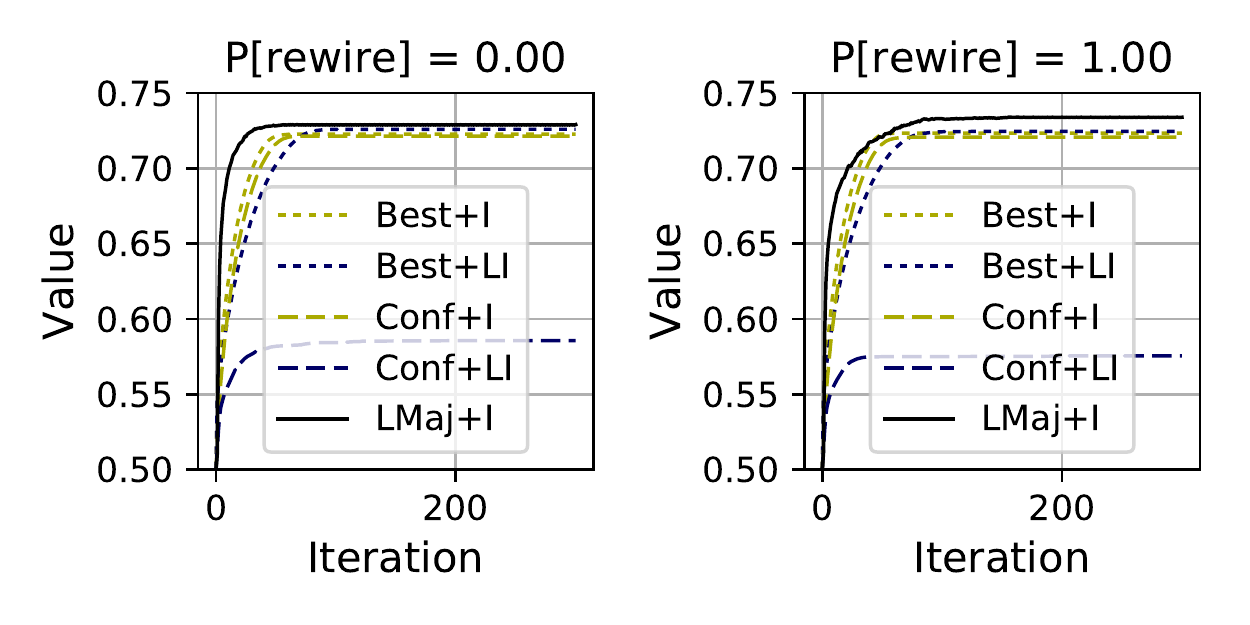}
\caption{
Mean agent solution value over time, averaged over 100 trials.
Strategies are defined in Table \ref{tab:strat}.
\label{fig:val-iter}
}
\end{figure}

\begin{table}
\small
\centering
\begin{tabular}{lll}
Strategy & Performance & Efficiency\\
\hline
Best+I  & 0.722 $\pm$ 0.001 & 0.0221 $\pm$ 0.0003 \\
Conf+I  & 0.721 $\pm$ 0.001 & 0.0174 $\pm$ 0.0002 \\
Best+LI & 0.726 $\pm$ 0.001 & 0.0131 $\pm$ 0.0002 \\
Conf+LI & 0.586 $\pm$ 0.001 & 0.030 $\pm$ 0.001 \\
LMaj+LI & 0.729 $\pm$ 0.001 & 0.046 $\pm$ 0.002 \\
\hline
\end{tabular}
\caption{
Simulated Performance and Efficiency.
Results shown for 100 trials with P[rewire] = 0.
Strategies are defined in Table \ref{tab:strat}.
Local strategies are less efficient than their non-local counterparts.
Local best-neighbor out-performs global,
while local conformity is the worst performer in all cases.
The local majority strategy is both most efficient and most performant.
\label{tab:sim-eff-perf}
}
\end{table}

\begin{table}
\small
\centering
\begin{tabular}{lll}
Strategy & Perf. Std. Coeff. & Eff. Std. Coeff.\\
\hline
Best+I  & -4.2$\times{10^{-5}}$ & \+4.1$\times{10^{-5}}$ \\
Conf+I  & \+2.7$\times{10^{-5}}$ & \+9.4$\times{10^{-5}}$ \\
Best+LI & -9.6$\times{10^{-4\;**}}$ & \+7.7$\times{10^{-5}}$ \\
Conf+LI & -1.5$\times{10^{-3\;***}}$ & \+8.7$\times{10^{-5\;*}}$ \\
LMaj+LI & \+1.2$\times{10^{-4\;**}}$ & -0.038$^{***}$ \\
\hline
\end{tabular}
\begin{tablenotes}
\item \centering * $p < 0.05$. ** $p < 0.01$. *** $p < 0.001$.
\end{tablenotes}
\caption{
Degree regression coefficients for simulations.
\label{tab:sim-perf}
}
\end{table}

\begin{figure*}
\includegraphics[width=6.67in,height=1.33in]{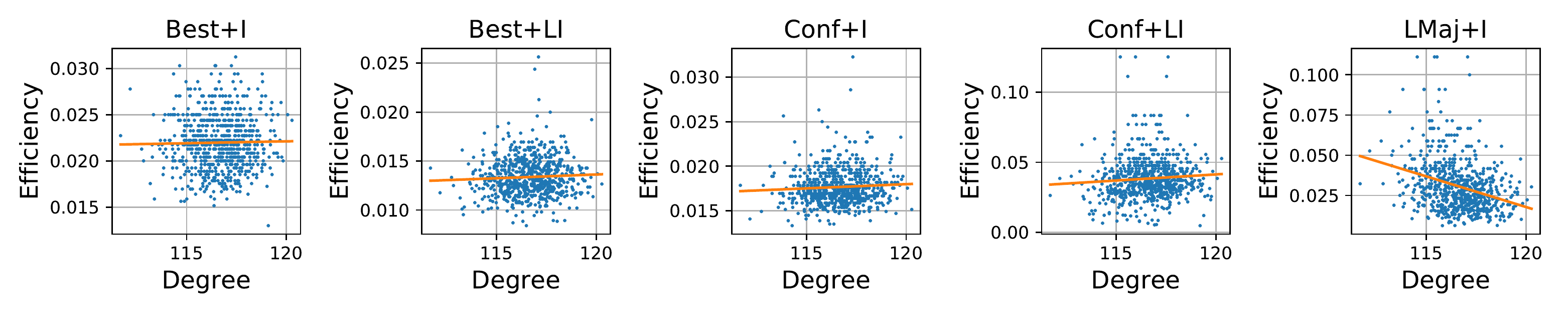}
\includegraphics[width=6.67in,height=1.33in]{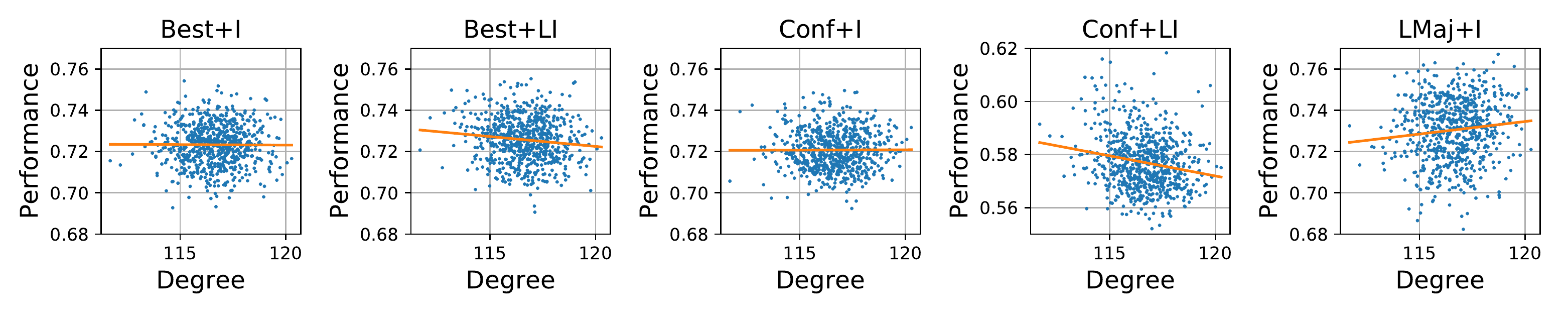}
\caption{
Efficiency and Performance of social learning strategies vs. mean network degree.
Each point represents a single trial of 300 iterations.
Strategies are defined in Table \ref{tab:strat}.
The local best-neighbor strategy shows
decreased performance at high degree,
with no significant change in efficiency.
Local conformity shows decreased performance and increased efficiency at high
degree.
Local majority shows the opposite behavior:
increased performance and decreased efficiency at high degree,
with the efficiency showing the largest effect size of all strategies.
\label{fig:deg-eff-perf}
}
\end{figure*}

Figure \ref{fig:val-iter} shows how agents' solutions improve after repeated applications of
different learning strategies and rewiring values.
Each curve represents an average over 100 trials, each with 250 agents.
The mean performance and efficiency are reported in Table \ref{tab:sim-eff-perf}.
For all rewiring values, local strategies are less efficient and more performant
than their non-local counterparts.
For the best-neighbor strategy, local outperforms global.
Local conformity performs notably worse than all others.
Local majority is both more efficient and more performant than others,
with its performance increasing with higher rewiring.
This implies that,
at least in a simple collaboration model,
performance and efficiency can be simultaneously increased.
Furthermore, performance and efficiency are potentially affected by both the choice of
learning strategy and the average degree of the agents' social network.

The effects of degree on performance and efficiency are shown in Figure \ref{fig:deg-eff-perf} and
Table \ref{tab:sim-perf}.
For non-local versions of both conformity and best-neighbor strategy,
there is no significant effect of degree on performance or efficiency.
The local best-neighbor strategy shows reduced performance with increasing degree,
but no change in efficiency.
Local conformity and local majority show opposite behavior as degree increases:
with local conformity gaining efficiency at the expense of performance,
while local majority increases in performance and decreases in efficiency.
The largest effect size is achieved for efficiency in the local majority simulation,
which is consistent with the efficiency behavior observed in WikiProjects.
However, the performance behavior for local majority is opposite that observed on
Wikipedia.
These agent-based models confirm that network degree has the potential to influence
the performance and efficiency of collaborations.
Furthermore, this influence can be drastically different depending on the strategies
used by collaborators.

\section{Discussion}
\label{sec:discuss}

While existing research into the role of network structure in collaboration
has focused on numerical simulations and lab experiments,
analysis of large real-world systems is an important next step.
Our empirical analysis contributes several findings towards a better
understanding of large, decentralized, real-world collaboration.
We observe several results consistent with previous work:
a trade-off between performance and efficiency
\cite{mason_propagation_2008,grim_scientific_2013},
higher performance for shorter path lengths in a conformity setting
\cite{barkoczi_social_2016},
and a reduction in performance with increased structural inequality
\cite{kearns_experiments_2012}.
By using real-world networks, we were also able to analyze network properties independently.
While most existing work has focused on the importance of path length,
our findings suggest degree distribution may be just as, or more, important.
The association of low degree with both high performance and high efficiency is compelling,
as it sidesteps the usual trade-off between performance and efficiency.
In low-degree networks, agents have more repeated interactions with smaller groups of collaborators, suggesting that small team sizes could be beneficial for large collaborations.
Similarly, the observation that performance is higher in projects with less structural inequality suggests that,
if the challenges of egalitarian organizing are overcome,
decentralized collaborations may produce better outcomes than those with centralized, top-down structures.

Our agent-based models offer a several insights.
We observe degree-dependent performance and efficiency
for both local conformity and local majority strategies.
However, these two strategies have opposite degree dependence,
suggesting that different strategies may be preferable for high-degree and low-degree
networks.
Our local majority strategy,
designed to satisfy several properties found in real-world collaborations,
shows the strongest effects on performance and efficiency as network degree changes.
For the local majority strategy,
the relationship between degree and efficiency is consistent with our
empirical observations on Wikipedia,
suggesting one possible mechanism underlying that efficiency dependence.
However, the performance dependence of this strategy
is opposite that observed on Wikipedia,
suggesting that either the local majority strategy
is incompatible with actual behavior on Wikipedia or
that other factors outweigh the contribution of mean degree.

Our work has several limitations.
Our empirical analysis is purely correlative and cannot be used to draw
conclusions about the causal influence of network structure on collaboration.
However, the consistency of our results with other lab-based and numerical studies
suggests that the causal link is worthy of further study.
Similarly, our study focuses entirely on a single online community,
and while the results are suggestive, they do not necessarily generalize.
We have focused on structure, ignoring content-related variables.
For simplicity, we have assumed unweighted edges and
measured work by revision counts rather than bytes changed.

Our work suggests several directions for future work.
Is the correlation between network structure, performance, and efficiency causal?
A time-dependent analysis of our data could offer insight.
Are similar relationships observed in other large-scale collaborations?
Does varying degree independently of path length influence
performance and efficiency in a controlled lab setting?

A better understanding of the relationship between network structure and collaboration
outcomes has practical applications.
Online communities using recommender systems could make recommendations guided by desirable network properties.
Similarly, network structure could be used to identify under-performing
groups in need of an intervention.
The relationship between network structure and learning strategy suggests that
behaviors interact with network structure,
which could be used to encourage behaviors complementary to existing network
structure.

\section{Conclusion}
\label{sec:conclusion}

In this paper, we have described the relationship between the structural properties of WikiProject
coeditor networks, their performance, and their efficiency.
As in other studies, we see a trade-off between performance and efficiency.
However, 
some properties, such as low degree, are associated with
both higher performance and higher efficiency.
We also find that the correlations between path length and performance are consistent with a conformity-based
social learning strategy, but not a greedy best-neighbor strategy.
We observe improved performance in more decentralized projects,
as has been seen in small-scale lab experiments.
We have also proposed a novel local majority learning strategy that is more realistic,
more efficient, and higher performance than existing strategies.
While most previous social learning simulations focus on path length,
we observe degree-dependent performance and efficiency
in both the local majority strategy
and a localized version of the conformity strategy.
We find that the direction of that dependence varies with the
specific strategy being used.
While additional work is needed to determine causal relationships
and the generalizability of our results,
we have shown evidence that several phenomena predicted by numerical
and small-scale lab experiments are present in a large,
real-world collaboration.
Our results suggest that the success of large-scale collaborations may be aided
by greater decentralization,
consensus or conformity-based decision-making,
and more tightly-knit collaborations between smaller teams.

\section{Acknowledgments}
Thanks to
Ceren Budak, Scott E. Page, Yan Chen, Tanya Rosenblat,
anonymous reviewers,
and the attendees of the May 25, 2017 MIT Center for Civic Media lab meeting
and the Berkman-Klein Center Cooperation Working Group for helpful feedback;
and to Danielle Livneh and Karthik Ramanathan for help collecting the data.
This research was partly supported by the National Science Foundation under Grant No. IIS-1617820.

\bibliographystyle{plain}
\bibliography{wp-eff-perf}

\end{document}